\documentclass[a4paper,aps,twocolumn,groupedaddress,showkeys,showpacs,floats,floatats,floatfix]{revtex4-1}

\usepackage{graphicx}  
\usepackage{graphics}
\usepackage{epsfig}
\usepackage{latexsym}
\usepackage{mathrsfs}
\usepackage{amssymb}
\usepackage{amssymb}
\usepackage{amsmath}
\usepackage{amscd}
\usepackage{color}
\usepackage{verbatim}
\begin{document}
\title{Three unequal masses on a ring and soft triangular billiards} 
\author{H.~A.~Oliveira}
\author{G.~A.~Emidio}
\author{M.~W.~Beims}
\email[E-mail address:~]{mbeims@fisica.ufpr.br}
\affiliation{Universidade Tecnol\'ogica Federal do Paran\'a,
         87301-006 Campo Mour\~ao, Brazil}
\affiliation{Departamento de F\'\i sica, Universidade Federal do Paran\'a,
         81531-990 Curitiba, Brazil}
\date{\today}

\begin{abstract}
The dynamics of three soft interacting particles on a ring is shown to 
correspond to the motion of one particle inside a soft triangular 
billiard. The dynamics inside the soft billiard depends only on the 
{\it masses ratio} between particles and {\it softness ratio}
of the particles interaction. The transition from soft to hard interaction 
can be appropriately explored using potentials for which the corresponding 
equations of motion are well defined in the hard wall limit. Numerical 
examples are shown for the soft Toda-like interaction and the error 
function.
\end{abstract}



\maketitle

{\bf Hard walls in billiard systems were extensively modelled by 
$\delta$-potentials since they allow for simple analytical 
relations of velocities and angles before and after the 
collisions with the walls. However, equations of motion are not
well defined at the collision point.
To analyze the transition to soft walls, which are more realistic,
it is essential to have well defined equations of motion since in
general no simple analytical solutions are obtained. The present
work suggests that appropriated soft walls are those for which the 
forces, not the potentials, become $\delta$-functions in the limit 
of hard walls. This allows for better numerical investigation
of the soft-hard transition. A general scaled Hamiltonian is derived 
for three unequal masses interacting particles on a frictionless 
ring, which nicely describes this transition and shows that the 
dynamics occurs inside a soft triangle billiard.}

\section{Introduction}
\label{Introduction}

Although physically realizable potentials are inherently soft, theoretical
models usually focus on hard potentials since they allow for analytical 
solutions. Billiard systems with hard walls are one example of such 
theoretical models and have been extensively study and well understood. 
For example, it is well know \cite{casati99,mittag96,casati97} that the 
motion of three particles on a frictionless ring with point-like interactions
is equivalent 
to the motion of one particle moving freely inside the triangular billiard 
which hard  walls. In such description the hard-walls potentials are 
represented by $\delta$-functions which makes the collisions with the 
walls very simple, and analytical results can be obtained for the real 
dynamics and also in the tangent space. However, not much has been done 
in the description of the transition from hard to soft interactions. Soft 
walls apparently do not destroy trajectories found in the hard-wall limit 
\cite{ana} and induce the appearance of regular islands in phase space 
\cite{turaev98,turaev99,turaev03,hercules08}. 
Some works about the dynamics in soft billiard have to be mentioned, 
one particle motion in an atom-optic billiard \cite{kaplan1,kaplan2}, 
quantum conductance in the soft wall microwave billiard~\cite{weingaertner},
the effect of wall roughness in granular Poiseuille flow~\cite{alam} and how 
the confinement of the equilibrium hard-sphere fluid to restrictive one- 
and two-dimensional channels with soft interacting walls modifies its 
structure, dynamics, and entropy~\cite{mittal06,mittal07} and  nonlinear 
dynamics in general \cite{donnay}.

There is a key issue for an appropriate description of soft to hard 
transitions in the context of billiards. Usually the hard walls are 
modelled by $\delta$-functions which allow for a simple description 
of the collision processes, since variables (angle and velocity) 
before and after the collisions with the walls can be given explicitly.
However, the corresponding equations of motion are not well defined. 
In the case of soft walls the variables before and after the collisions 
cannot be given in general explicitly, and equations of motion must be 
solved numerically. Thus, in order to study continuously the transition 
from soft to hard walls, equations of motions must be well defined in
all cases.  A first contribution \cite{hercules08} in this direction 
analyzed the dynamics of two interacting 
particles inside a 1D billiard with soft walls, where the soft walls were
modelled by the error function, which describes continuously and 
correctly the mentioned transition. The purpose of the present 
work is to generalize these results to the case of three soft interacting 
particles on a ring, and analyze the correct transition to point-like 
collisions. 
A general scaled Hamiltonian is derived which nicely describes this 
transition and also shows that the dynamics occurs inside a soft 
triangle billiard. The dynamics depends only on the {\it masses ratio} 
between particles and {\it softness ratio} of the interaction.

The paper is organized as follows. While in Section~\ref{triangle} 
the hard wall triangle billiard is revisited, Section \ref{soft-triangle} 
presents the general soft Hamiltonian system for three interacting 
particles, our main analytical result. In Section \ref{TODA} we apply 
the soft Hamiltonian to the Toda model with unequal masses. Section 
\ref{error} presents the example of two suitable potential which could 
be used to appropriately describe the soft to hard transition for three 
particles. Finally Section \ref{conclusions} summarizes our main results.

\section{Revisiting the hard triangular billiard}
\label{triangle}

It is well know \cite{casati99,mittag96,casati97} that the motion of three 
particles on a frictionless ring with point-like interactions is 
equi\-va\-lent 
to one particle moving freely inside the triangular billiard with angles

\begin{equation}
\tan{\alpha}=\sqrt{\frac{m_2M}{m_1m_3}},\
\tan{\beta}=\sqrt{\frac{m_1M}{m_2m_3}}, \
\tan{\eta}=\sqrt{\frac{m_3M}{m_1m_2}},
\nonumber
\label{angles}
\end{equation}
and colliding elastically with the sides of the triangle. Here
$M=m_1+m_2+m_3$ and $m_i$ is the mass of the particles ($i=1,2,3$).
The Hamiltonian can be written as 

\begin{equation}
H_B= K_M+ \delta(q_2-q_1)+\delta(q_3-q_2)+\delta(q_1-q_3+L).
\label{Hdelta}
\end{equation}
where $K_M=p_1^2/2m_1+p_2^2/2m_2+p_3^2/2m_3$ is the mass dependent 
kinetic energy. The collisions occur at $q_1=q_2, q_2=q_3$ and $q_1=q_3+L$ 
where $L$ is the circumference of the ring.
The point-like collision between particles $1$ and $2$ defines one 
side of the triangle at $q_1-q_2=0$, and the collision of 
these particles with particle $3$ defines the other two sides 
of the same triangle. For $m_3\rightarrow\infty$ ($\eta=\pi/2$) 
we get the right triangular billiard which corresponds to the 
motion of two particles $m_1$ and $m_2$ moving inside the 1D
box with hard walls.  In this case the interaction 
between particles $1$ and $2$ is the point-like collision and the 
fixed particle $3$ plays the role of a 1D hard-wall. In such systems 
the Lyapunov exponent is zero \cite{denisov05,grassberger02} and 
the whole dynamics can be monitored by changing the angles of 
the triangle billiard \cite{casati99}. It was shown \cite{cesar1}
that the Yukawa interaction between particles $1$ and $2$ is 
enough to generate positive Lyapunov exponents. It is also worth 
to mention it is possible \cite{cesar2} to relate the linear 
instability inside the triangular billiard with the Lyapunov 
exponents from quadratic irrational numbers, which are related 
to the angles of the triangle, and thus to the masses ratio.

\section{The soft triangular billiard}
\label{soft-triangle}

As observed in Hamiltonian (\ref{Hdelta}), the collisions with the 
fixed hard-wall can be represented by $\delta$-potentials. However, 
the corresponding equations  of motion are not well defined. Therefore, 
to describe analytically the transition to hard walls, we include soft 
interactions between particles which, in a given limit, are expected 
to describe the collisions with the hard-walls. Let us start with the 
Hamiltonian of the three particles on a frictionless ring given by 

{\begin{eqnarray}
H_B&=&K_M+V_{12}\left(\frac{q_2-q_1}{\sigma_{12}}\right)
\nonumber
\\
&+& V_{23}\left(\frac{q_3-q_2}{\sigma_{23}}\right)
+V_{31}\left(\frac{q_1-q_3+L}{\sigma_{31}}\right),
\label{Hsoft}
\end{eqnarray} 
where $\sigma_{ij}$ ($i,j=1,2,3$) defines the softness} of each pairwise 
interaction. It is assumed that the interaction potential $V_{ij}$ between 
the particles depends only on the re\-la\-tive position between them and
that for $\sigma_{ij}\to 0$ a $\delta$-like function is obtained for the 
corresponding {\it force}, and {\it not} for the potential. In addition, bounded
motion is expected below a certain energy which will be specified later.  
Before going into details about the appropriated potential $V_{ij}$ which 
could be used, we rewrite Hamiltonian (\ref{Hsoft}). Using the orthogonal
transformation \cite{casati99}:

\begin{eqnarray}
q_1 &=& -\sqrt{\frac{m_3}{(m_1+m_2)M}}x - \frac{1}{m_1}
         \sqrt{\frac{m_1m_2}{(m_1+m_2)}}y+\frac{z}{\sqrt{M}},\cr
    & & \cr
    & & \cr
q_2 &=& -\sqrt{\frac{m_3}{(m_1+m_2)M}}x +  
        \frac{1}{m_2}\sqrt{\frac{m_1 m_2}{(m_1+m_2)}}y+\frac{z}{\sqrt{M}},\cr
   & & \cr
   & & \cr
q_3 &=&  \sqrt{\frac{(m_1+m_2)}{m_3M}}x + \frac{z}{\sqrt{M}},
\label{trans}
\end{eqnarray}
for the three particles on a ring, and the linear transformations 
$x^{\prime}=\beta x$,  $y^{\prime}=\beta y$, $z^{\prime}=\beta z$, 
$d\tau/dt=\beta=\frac{1}{\sigma_{12}\sqrt{\mu_{12}}}$,
the final scaled Hamiltonian reads (without primes)

\begin{equation}
  H_B = K + {V_{12}(  y)}+ 
 {V_{23}(  x,  y)}+ V_{31}(  x,   y),
\label{HBS}
\end{equation}
where $K={  p}_x^2/2+{  p}_y^2/2+{  p}_z^2/2$
is the mass independent kinetic energy and
{
\begin{eqnarray}
 V_{12}(  y) &= &   V_{12}\left(-  y\right), 
\cr
   & & \cr 
V_{23}(  x,   y)   &=&   V_{23}
\left[\frac{\sigma_{12}}{\sigma_{23}}(a  y-b  
   x)\right],
\\
   & & \cr
V_{31}(  x,   y)&=&
V_{31}\left[\frac{\sigma_{12}}{\sigma_{31}}
(c  y+b  x) - \frac{L}{\sigma_{31}} \right].
\nonumber
\label{VBS}
\end{eqnarray}} 
Here $\mu_{21}=m_1m_2/(m_1+m_2)=m_{2}/(1+\gamma_{21})$ is the reduced 
mass between particles $1$ and $2$, $\gamma_{ij}=m_i/m_j$ is the 
mass ratio between particles $i$ and $j$, and 
$a=\frac{1}{(1+\gamma_{21})}$,
$b=a\sqrt{\frac{\gamma_{21}+\gamma_{31}+1}{\gamma_{32}}}$,
$c=a\gamma_{21}$. Since $\beta$ scales all coordinates, they are 
given in terms of the smoothness of the interaction between 
particles $1$ and $2$ and the reduced mass $\mu_{12}$.

\begin{figure}[htb]
 \unitlength 1mm
 \includegraphics*[width=8.0cm,angle=0]{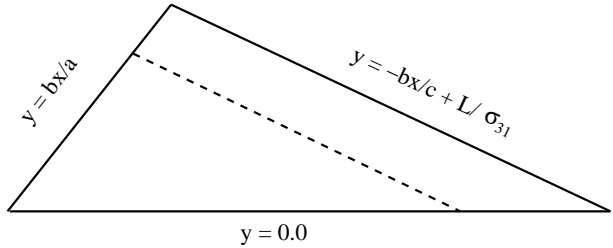}
 \caption{Scaled soft triangle. Dashed line represent changes of
the wall position when the softness parameter $L/\sigma_{31}$ varies.}
  \label{triangleFIG}
  \end{figure}

The Hamiltonian (\ref{HBS}) represents one particle with scaled 
mass $  m=1$ inside a triangular potential with three {\it soft} 
walls located at $y=0,y=\frac{b}{a}x$ and 
$  y=-\frac{b}{c}  x+L/\sigma_{31}$ (See Fig.~\ref{triangleFIG}).
The smoothness of the walls will depend not only on the form of the 
potentials $V_{12}(  y),  V_{23}(  x,  y), V_{31}(  x,  y)$ and the 
smoothness parameters $\sigma_{12}, \sigma_{23},\sigma_{31}$, but also on 
$a,b,c$  which depend only on the masses {\it ratios}.

The Hamiltonian (\ref{HBS}) is quite interesting since some general physical 
situations can be observed without choosing a specific form for the 
interaction potentials: 

\noindent
(i) The $z$ dependence
dissapeared due to the translational symmetry of (\ref{Hsoft}). It implies the
conservation of the total linear momentum.\\
\noindent (ii) the coupling between center of mass ($x$) and relative 
($y$) coordinates appears only in the interaction potentials 
$V_{31}(x,y)$ and $V_{23}(x, y)$. Thus only these 
potentials may generate a chaotic dynamics inside the billiard. When 
$ V_{31}= V_{23}=0$  the Hamiltonian (\ref{HBS}) 
is separable and thus integrable, as expected.\\
\noindent (iii) Softness affects simultaneously the center of mass and relative
coordinates since the ratios $ \sigma_{12}/\sigma_{23},  \sigma_{12}/\sigma_{31}$
multiply both variables ($  x, y$). These quantities does not 
change the internal angles of the soft billiard.\\
\noindent (iv) Changes in the masses ratio will modify the parameters $a,b$ 
and $c$ separately, so that center of mass and relative coordinates will 
vary separately. Only these quantities can change the internal angles of 
the billiard.\\
\noindent (iv) Keeping the masses ratio constant, the softness inside 
the triangle billiard is tailored by the
{\it ratios} $ \sigma_{12}/\sigma_{23}$, $\sigma_{12}/\sigma_{31}$ and
$L/\sigma_{31}$. The hard wall limit is obtained by making 
$L/\sigma_{31}\to\infty$, i.~e.~the wall $V_{31}$ moves to the right 
maintaining its inclination and the triangle size increases more and more. 
See dashed lines in Fig.~\ref{triangleFIG}.\\

The Hamiltonian (\ref{HBS}) is quite general. Any interaction 
potential $V_{12}, V_{23}, V_{31}$ between particles can be used: 
Coulomb, Yukawa, Wood-Saxon, exponential etc. If the potential 
obeys the softness limits mentioned above, the properties (i)-(iv) 
should be valid for {\it any} of such in\-te\-rac\-tions. Just to 
exemplify this we discuss next the well known Toda potential 
\cite{toda70}.

\section{Toda potential}
\label{TODA}

\subsection{Equal masses}
\label{todae}

We consider the three-particle Toda lattice \cite{toda70}
whose potential is obtained by using $V_{12}=e^{(q_2-q_1)}, 
V_{23}=e^{(q_3-q_2)}, V_{31}=e^{(q_1-q_3)}$ with $L=0$ so that

\begin{equation}
H_{T}^e=K + e^{(q_2-q_1)}+e^{(q_3-q_2)}+e^{(q_1-q_3)},
\label{toda}
\end{equation}
and corresponds to three equal masses particles moving on a ring. 
The letter $e$ stands for equal masses.
Here the variables $q_1,q_2,q_3$ are the angles of particles related 
to the origin of the ring. In addition to the total energy and the linear 
momentum, this  problem is know to have a third non-trivial integral of
motion (For more details see \cite{lichtenberg92}). Using the 
transformation (\ref{trans}) for the particular case
of equal masses, the following two dimensional potential is obtained:

\begin{equation}
V_{T}^e=\left[e^{(2y+2\sqrt{3}x)}+e^{(2y-2\sqrt{3}x)}+e^{(-4y)}\right].
\label{toda2D}
\end{equation}
The equipotential lines for this potential look like from a soft 
triangle (See Fig.~(1.8) from \cite{lichtenberg92}). Therefore the 
analogy with the soft triangular billiard is evident. The additional 
isolated integral of motion is \cite{henon74}:
\begin{eqnarray}
I&= & 8p_x(p_x^2-3p_y^2)+(p_x+\sqrt{3}p_y)e^{2y-2\sqrt{3}x} \cr
& & \cr
& & -2p_x e^{-4y} + (p_x-\sqrt{3}p_y)e^{2y+2\sqrt{3}x}, 
\label{I}
\end{eqnarray}
which is not related to any obvious conservation law or symmetry. 
Potential (\ref{toda}) is a nice example of the soft 
triangular description from Eq.~(\ref{HBS}).

\subsection{Unequal masses}
\label{todau}

The unequal mass Toda problem was firstly studied numerically
\cite{casati75} for the case of two masses interacting in the 
Hamiltonian

\begin{equation}
H_{T}^u=K_M+ e^{-q_1}+e^{-(q_2-q_1)}+ e^{q_2}-3,
\label{casati}
\end{equation}
where $u$ denotes unequal masses.
They showed the transition to stochasticity when $m_1\ne m_2$ 
and confirmed the integrability for $m_1=m_2$. The three unequal 
masses particles was analyzed \cite{dorizzi84} for a free-end lattice
Hamiltonian $H_{T}^{free}=K_M+e^{\epsilon(q_1-q_2)} + e^{(q_2-q_3)}$.  They 
found that this problem is integrable when 
$m_1=\epsilon (2\epsilon-1)/(2-\epsilon),
m_2=2\epsilon-1 $ and $m_3=1$. The parameter $\epsilon$ must satisfy
$1/2<\epsilon<2$. The Hamiltonian $H_{T}^{free}$ describes a scattering 
problem and the integrability cannot be verified numerically. Compared 
to the potential (\ref{toda}), the system  $H_{T}^{free}$ does not 
consider the interaction bettwen particles $1$ and $3$ because it is 
along a lattice line. See also \cite{yoshida87} for more about the 
integrability of Toda kind lattices. The full problem of the Toda-like 
Hamiltonian with three unequal 
masses ($m_1=m_2=m_3=1$) was studied before \cite{bountis82} and has the 
form $H_{T}=K_M+e^{\delta(q_1-q_2)}+ e^{\epsilon(q_2-q_3)} + e^{(q_1-q_3)}$. Using 
the Painlev\'e property, they found the system to be integrable only for 
$m_1=m_2=\delta=\epsilon=1$, which is again the 
Toda case discussed above. 

Although it is not the purpose of the present work to study 
the Toda potential, it can be written in the form 
$H=K_M+V_{T}^u$ where

\begin{equation}
V_{T}^u=e^{-\frac{1}{\sigma_{13}}(q_1-q_3)}+e^{-\frac{1}{\sigma_{21}}(q_2-q_1)}+
    e^{-\frac{1}{\sigma_{32}}(q_3-q_2)},
\label{todasoft}
\end{equation}
which now includes the the softness parameters $\sigma_{ij} $ between 
each pair ($i,j$) interaction. Using Eqs.~(\ref{trans}) in the potential 
(\ref{todasoft}) we obtain the two dimensional version

\begin{eqnarray}
  V_T^u  =  e^{\left(-  y\right)}+
e^{\left[\frac{\sigma_{12}}{\sigma_{23}}(a  y-b  
   x)\right]}+
e^{\left[-\frac{\sigma_{12}}{\sigma_{13}}(c 
  y+b  x)\right]}.
\label{todasoft2D} 
\end{eqnarray}
The whole mass dependence is now inside the parameters 
from the potential $  V_T^u$. The non trivial invariant (\ref{I}) 
does not exist anymore for the case of unequal masses. This was 
checked by testing the condition $dI^u/dt=0$, where $I^u$ is the 
conservative quantity, similar to Eq.~(\ref{I}), but for the case 
of unequal masses.

The important point to mention here is that the integrability of the 
system (\ref{todasoft2D}) depends only on the {\it masses ratios} 
$m_2/m_1=m_3/m_1=m_3/m_2=1$ and the {\it softness ratios} 
$\sigma_{12}/\sigma_{23}=\sigma_{12}/\sigma_{13}=1$. This 
generalizes the result from \cite{bountis82}.

\begin{figure}[htb]
 \unitlength 1mm
 \includegraphics*[width=4.0cm,angle=0]{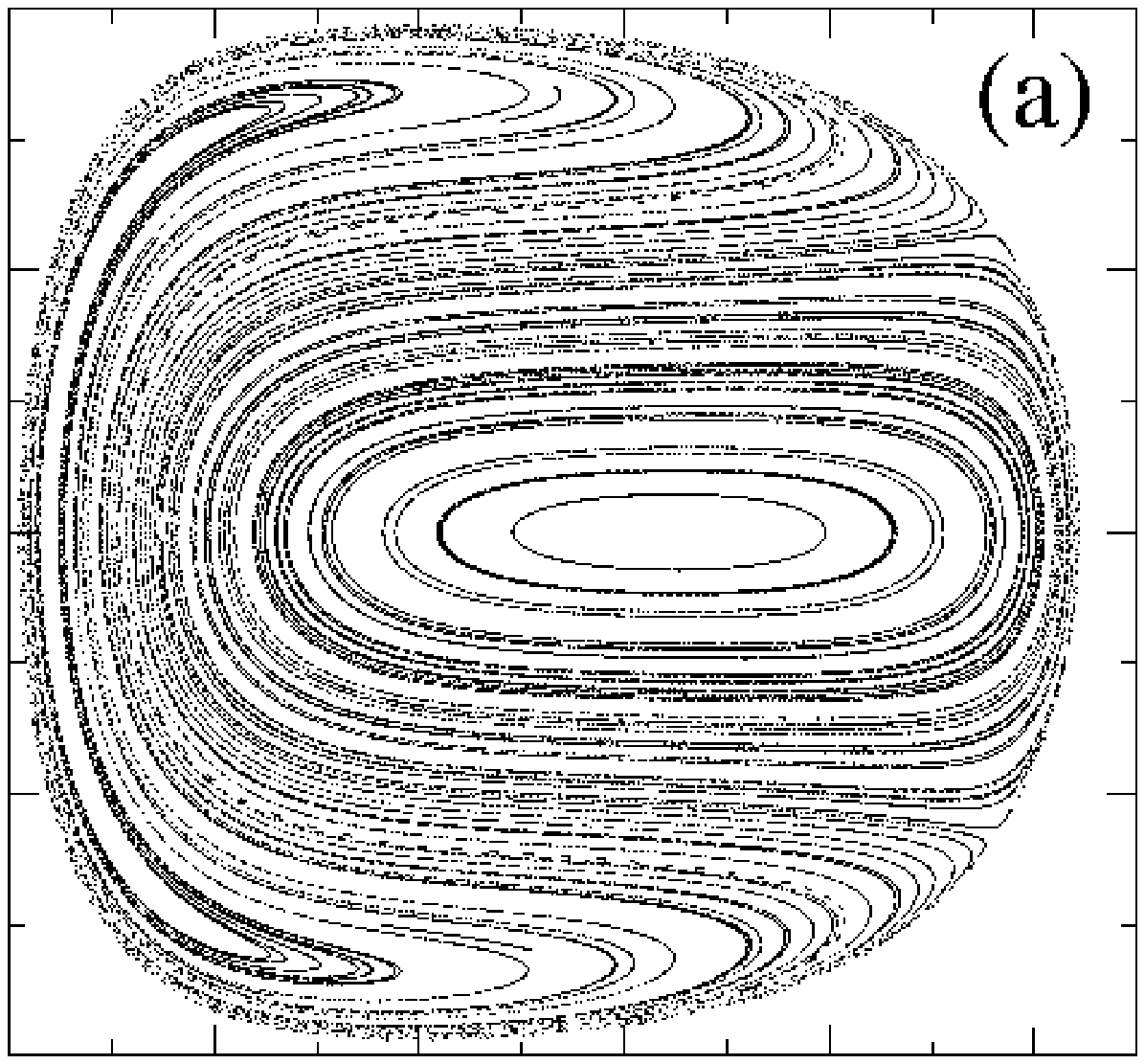}
 \includegraphics*[width=4.0cm,angle=0]{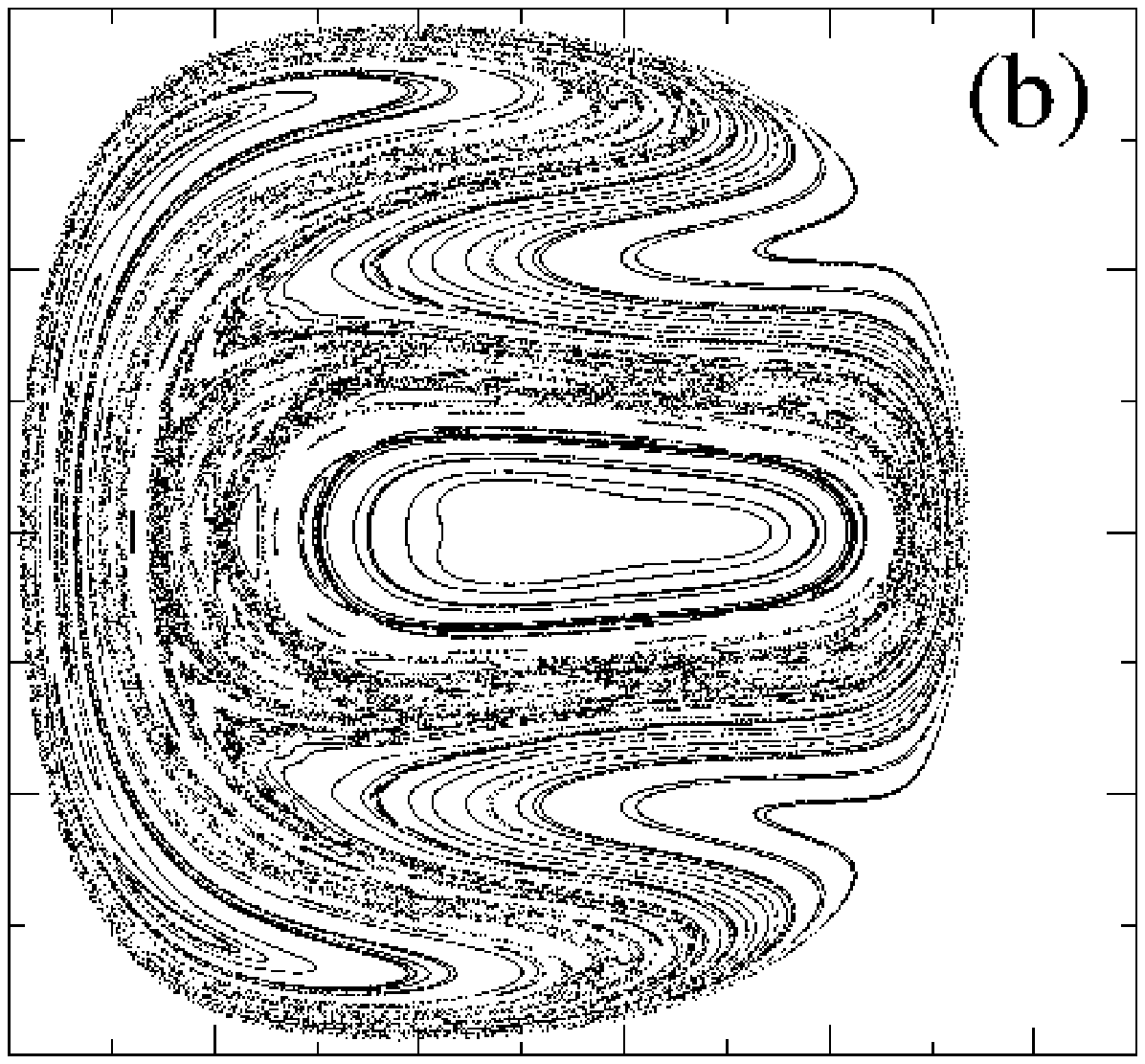}
 \includegraphics*[width=4.0cm,angle=0]{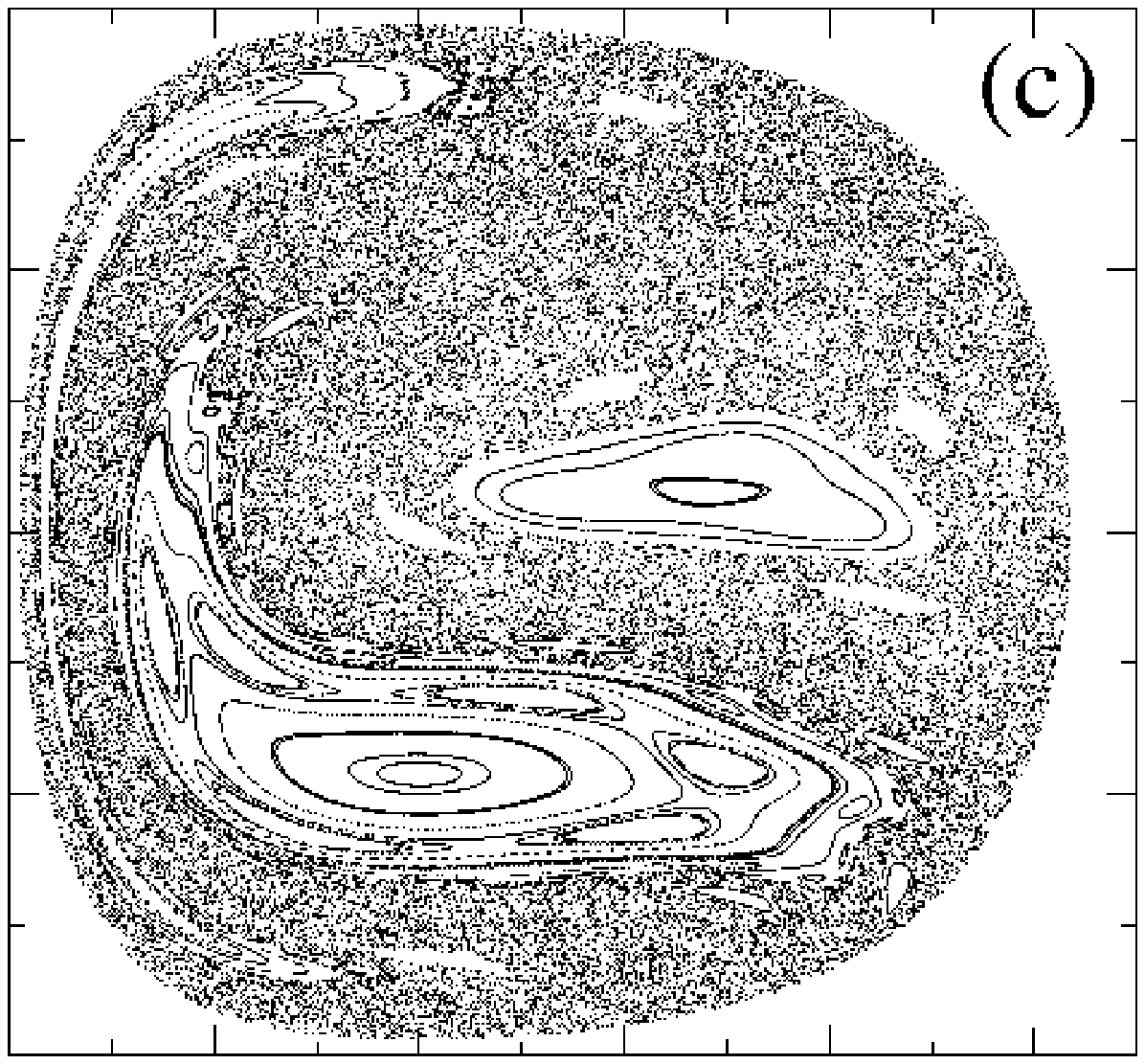}
 \includegraphics*[width=4.0cm,angle=0]{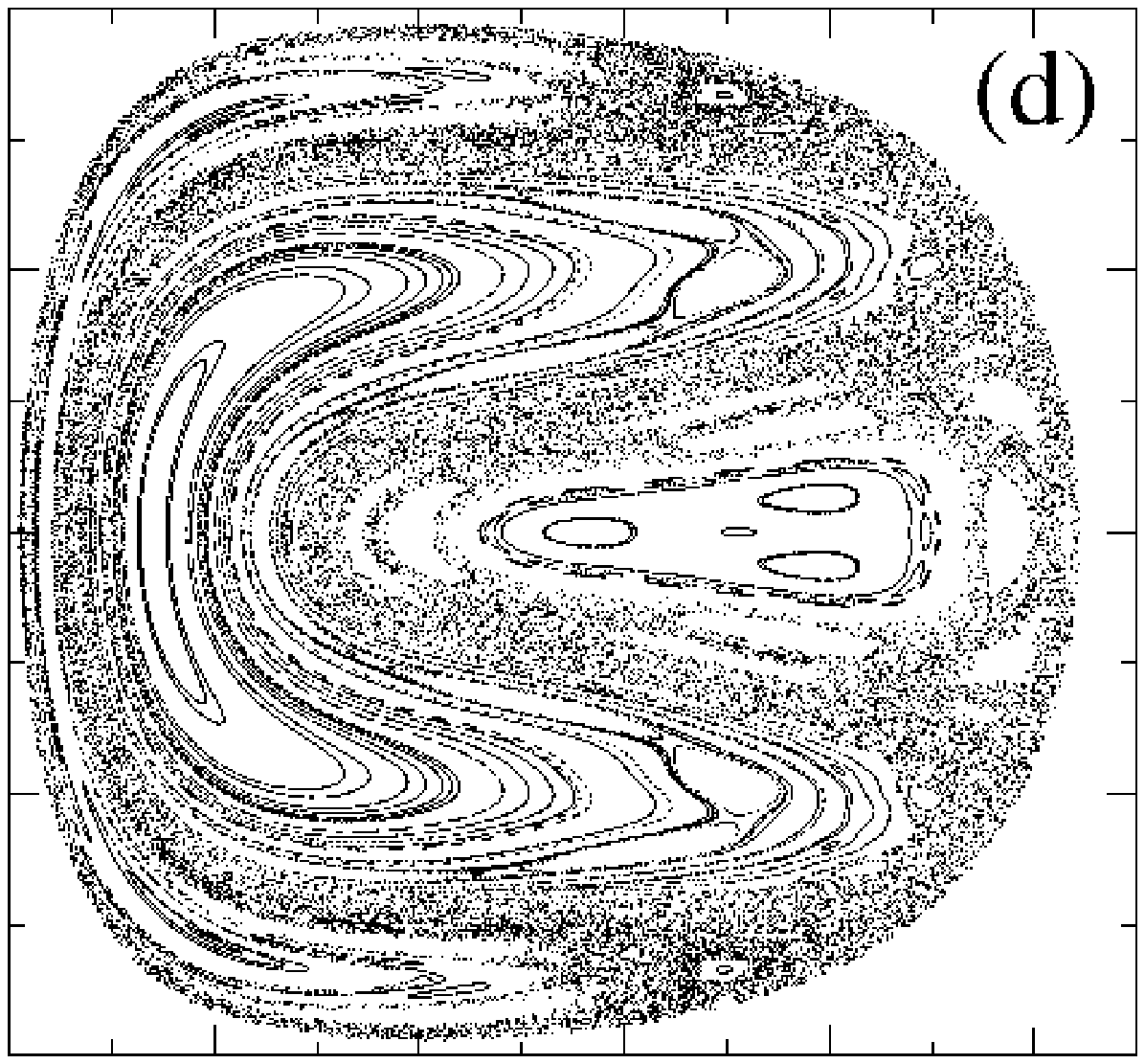}
 \caption{Poincar\'e Surfaces of Section 
($-4.0\le y \le 7.0,-10\le p_y\le 10$) for $E=50$ and (a) 
Integrable Toda case $\gamma_{21}=\gamma_{31}=\gamma_{32}=1.0$,  
$\sigma_{12}/\sigma_{23}=\sigma_{12}/\sigma_{13}=1.0$, 
(b) small softness asymmetry $\gamma_{21}=\gamma_{31}=\gamma_{32}=1.0$,  
$\sigma_{12}/\sigma_{23}=\sigma_{12}/\sigma_{13}=1.2$ and the
masses symmetric cases ($\sigma_{12}/\sigma_{23}=\sigma_{12}/\sigma_{13}=1.0$)
(c) $\gamma_{21}=\gamma_{31}=1.2, \gamma_{32}=1.0$ and
(d) $\gamma_{21}=1,\gamma_{31}=\gamma_{32}=1.2$.}
  \label{PSS1-3}
  \end{figure}

In a realistic problem, like quantum dots for example, the 
material of the boundaries has impurities so that the interaction 
softness between different particles may suffer small changes. 
In such cases variations of the softness {\it ratios} may destroy the 
integrability. This can be very nicely seen in Fig.~\ref{PSS1-3} 
where the Poincar\'e Surface of Section (PSS) is shown for small 
variations of the softness ratios. Figure \ref{PSS1-3}(a) is the 
integrable case  $m_2/m_1=m_3/m_1=m_3/m_2=1$ and 
$\sigma_{12}/\sigma_{23}=\sigma_{12}/\sigma_{13}=1$, while 
Fig.~\ref{PSS1-3}(b) 
we used $\sigma_{12}/\sigma_{23}=\sigma_{12}/\sigma_{13}=1.2$, keeping the
masses ratios equal one. A change in the dynamics is observed
and the chaotic motion becomes stronger.
It is also possible to  change the masses ratio keeping the softness 
ratios equal one, as can be see in Fig.~\ref{PSS1-3}(c) and (d). Strong 
chaotic behavior occurs when $\gamma_{21}=\gamma_{31}=1.2, \gamma_{32}=1$, 
but a not that large chaotic motion for 
$\gamma_{21}=1,\gamma_{31}=\gamma_{32}=1.2$ in Fig.~\ref{PSS1-3}(d).

\section{Soft suitable potentials}
\label{error}

\begin{figure}[htb]
 \unitlength 1mm
 \begin{center}
 \includegraphics*[width=4.2cm,angle=0]{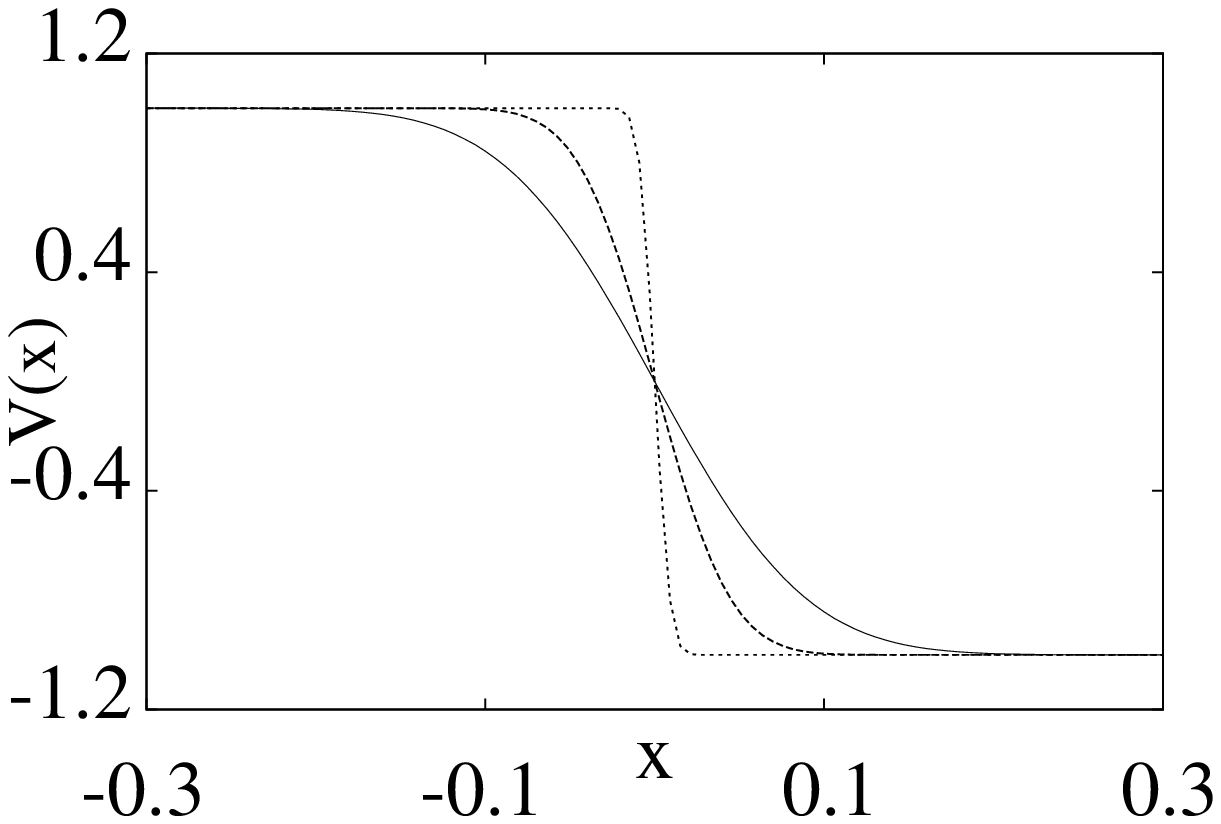}
 \includegraphics*[width=4.2cm,angle=0]{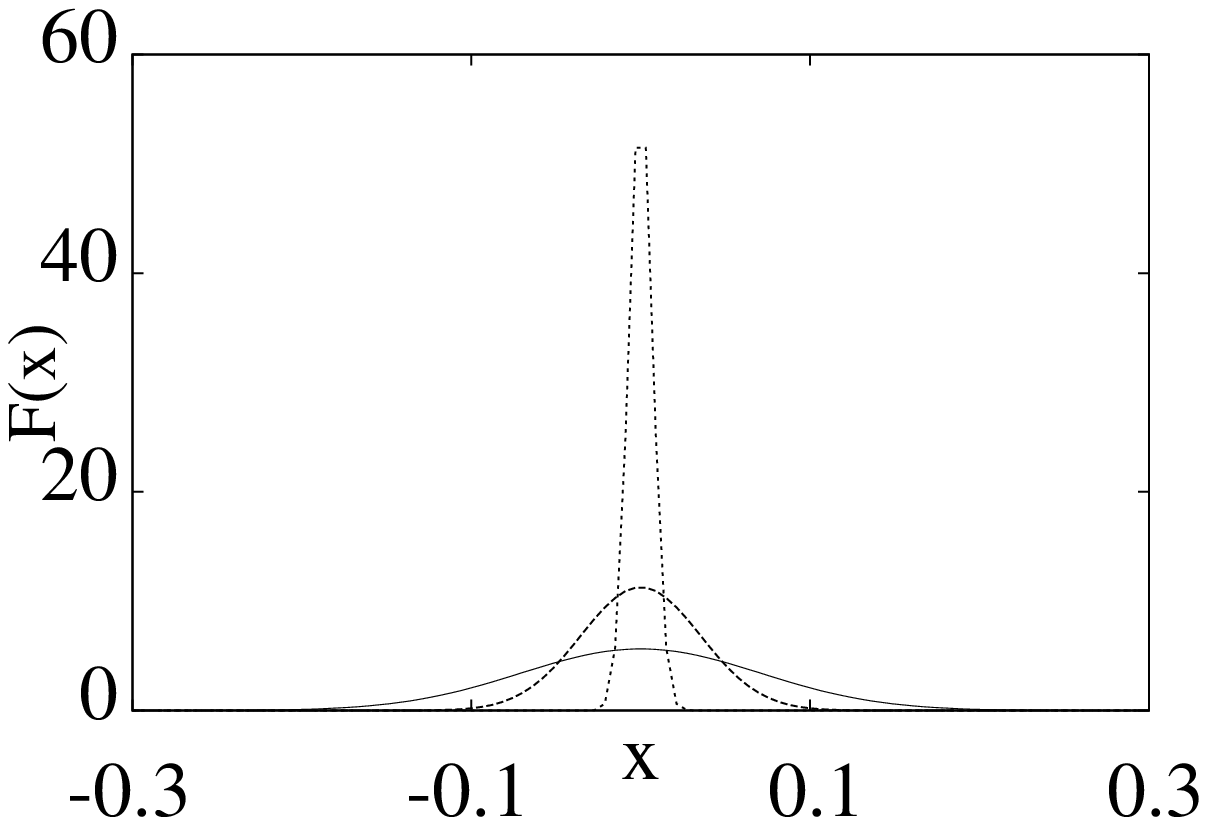}
 \leavevmode
 \end{center}
 \caption{(Color online) (a) Potential $V_{erf}(x)$ and 
(b) force $F_{erf}(x)$ for softness values $\epsilon=0.1$ 
(continuous line) $\epsilon=0.05$ (dashed) and 
$\epsilon=0.01$ (points).}
\label{pot-force}
  \end{figure}

To study the apropriate soft hard transition, in this work 
we assume potential functions which in limit of hard-walls 
are {\it not}  $\delta$-functions but the 
corresponding {\it forces} are. In this way the potential and 
forces are well defined for any softness of the potential.  
Consider, for example, the following three potentials and the 
cor\-responding forces:

\begin{eqnarray}
V_{exp}(x) & = & e^{-|x|/\epsilon},\qquad\qquad
\quad F_{exp}(x) = \frac{1}{\epsilon}e^{-|x|/\epsilon}, \cr
& & \cr
V_{erf}(x) & = & \frac{\sqrt{\pi}}{2}Erf{\left(\frac{x}{\epsilon}\right)},\qquad 
\,F_{erf}(x) = \frac{1}{\epsilon}e^{-x^2/\epsilon^2}, \cr
& & \cr
V_{arc}(x) & = & -Arctan{\left(\frac{x}{\epsilon}\right)},\qquad 
F_{arc}(x) = -\frac{\epsilon}{x^2+\epsilon^2}.\cr
\nonumber
\end{eqnarray}
In the hard wall limit $\epsilon\to 0$ the forces 
$F_{exp}(x),F_{erf}(x)$ and $F_{arc}(x)$  approach the well defined 
$\delta$-function. For example, Fig.~\ref{pot-force} shows the (a) 
potential $V_{erf}(x)$ and (b) force $F_{erf}(x)$ for distinct
softness values.
They show how the transition from soft to hard walls potentials
and forces are well behaved. There are obviously other functions
which in the limit $\epsilon\to 0$ behave like $\delta$-functions, 
but they are not appropriate for the purpose of the present work
which considers billiard walls. To avoid a too long paper, we just 
discuss the potentials $V_{exp}(x)$ and $V_{erf}(x)$. Results for 
$V_{arc}(x)$ are approximately similar (not shown here) to $V_{erf}(x)$.

Applying the potentials $V_{exp}(x)$ and $V_{erf}(x)$ in the 
Hamiltonian (\ref{HBS}) we obtain respectively the following 
systems

\begin{eqnarray}
H_B^{exp}&=& K + 2\left[e^{(-  y)}+
e^{\frac{\sigma_{12}}{\sigma_{23}}
\left (a  y-b  x\right)} + e^{\frac{\sigma_{12}}{\sigma_{31}}
\left (c  y+b  x -\frac{L}{\sigma_{31}}\right)}\right], 
\cr
& & \cr
  H_B^{erf}&=& K + Erf(-  y)+
Erf\left[\frac{\sigma_{12}}{\sigma_{23}}
(a  y-b  x)\right] \cr
& &\cr
&+& Erf\left[\frac{\sigma_{12}}{\sigma_{31}}
(c  y+b  x) -\frac{L}{\sigma_{31}}\right] + V_0, 
\label{three}
\end{eqnarray}
For $H_B^{erf} $ we added a constant potential $V_0=3.0$ so 
that both systems have the same total bounded energy 
$0.0\le  E_B\le 2.0$. For $E_B>2.0$ the dynamics is 
unbounded.

\subsection{The soft triangles}

In this section we show that the potential energy from
$H_B^{exp}$ and $H_B^{erf}$ have the form of a soft triangle. 
These potentials are shown,
 \begin{figure}[htb]
 \unitlength 1mm
 \begin{center}
 \includegraphics*[width=3.8cm,angle=0]{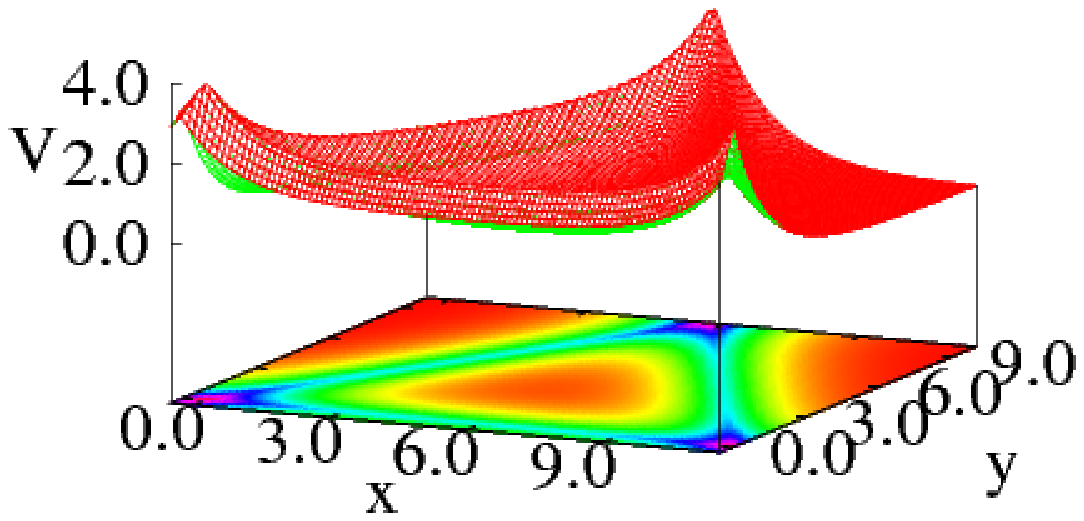}
 \includegraphics*[width=3.8cm,angle=0]{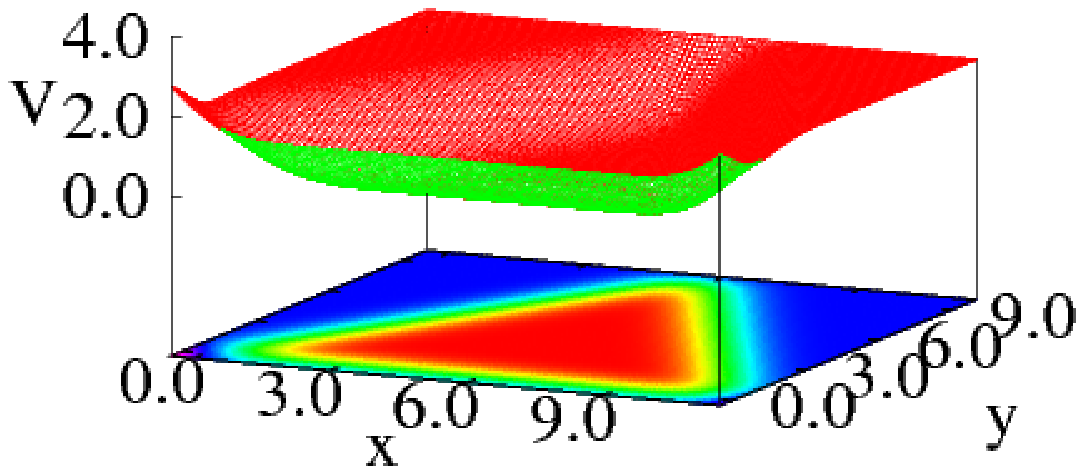}
 \includegraphics*[width=3.8cm,angle=0]{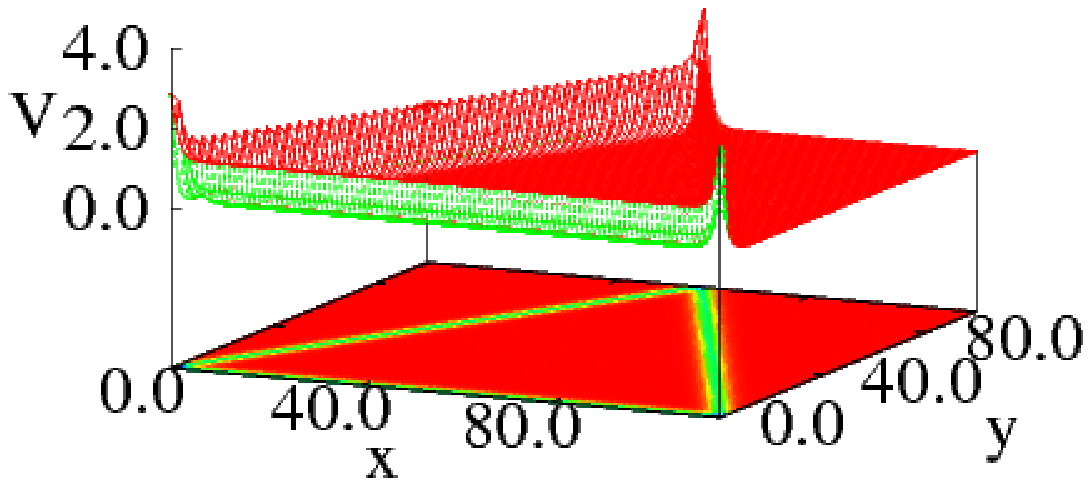}
 \includegraphics*[width=3.8cm,angle=0]{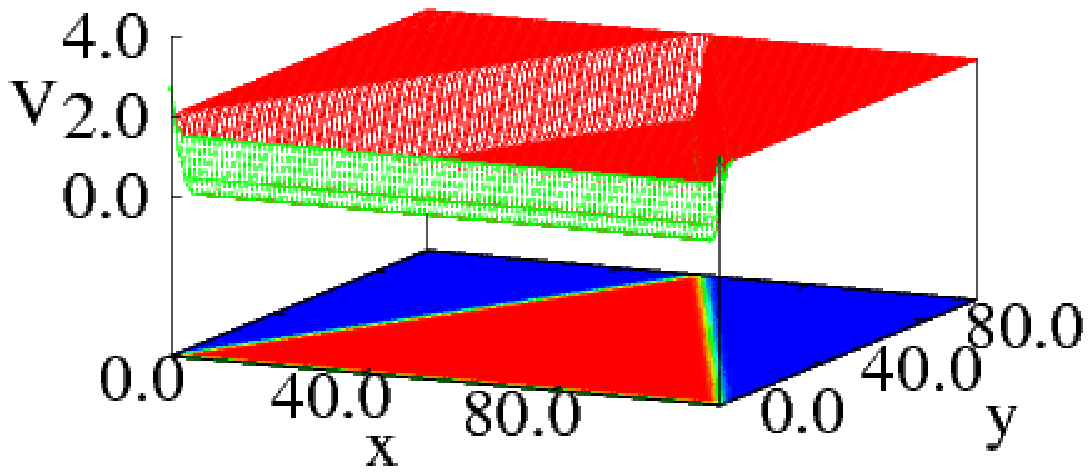}
 \end{center}
 \caption{(Color online) Potential energy from $H_B^{exp}$ (left)
and $H_B^{erf}$ (right) for the case of equal masses ratio and 
equal softness ratio. Top row for the soft case $L/\sigma_{31}=10$ 
and botton row for  $L/\sigma_{31}=100$ approaching the hard-wall 
case.}
  \label{potentials}
  \end{figure}
respectively in Fig.~\ref{potentials} for the case of equal softness
and equal masses ratio. Top row for $L/\sigma_{31}=10$ and bottom row
for $L/\sigma_{31}=100$.  First observation is that both potentials 
have a triangle billiard-like form (see projection on the $x,y$ 
plane) and that the walls are soft for $L/\sigma_{31}=10$. As this 
ratio increases to $L/\sigma_{31}=100$ the walls start to look more 
similar to hard-walls. For the purpose of clarity, in the left plots 
we used the modulus in the argument of the exponential potentials. 
Otherwise it would  not be possible to plot the coordinates in an 
adequate way to see the triangle form.

\begin{figure}[htb]
 \unitlength 1mm
 \begin{center}
 \includegraphics*[width=3.8cm,angle=0]{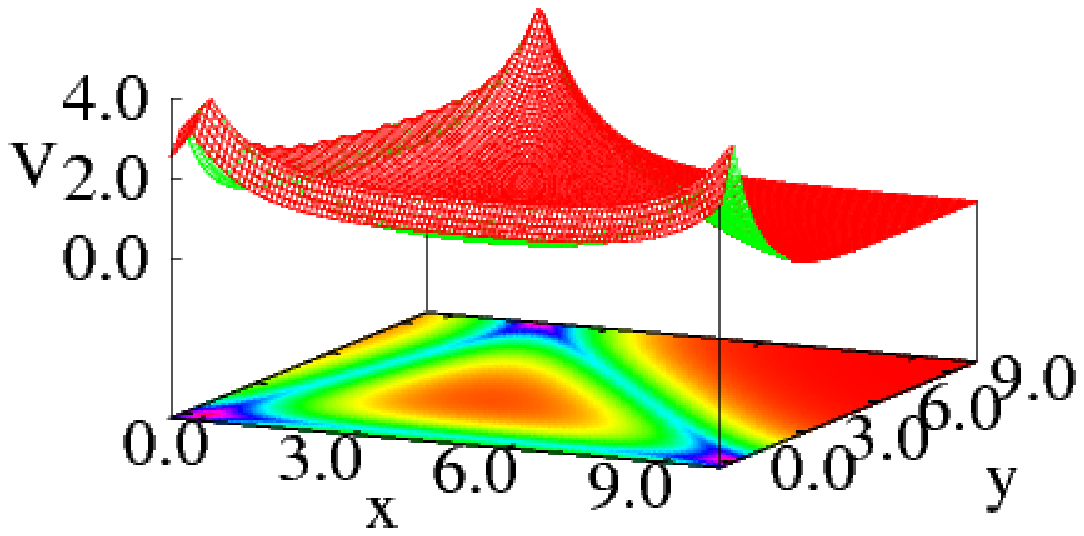}
 \includegraphics*[width=3.8cm,angle=0]{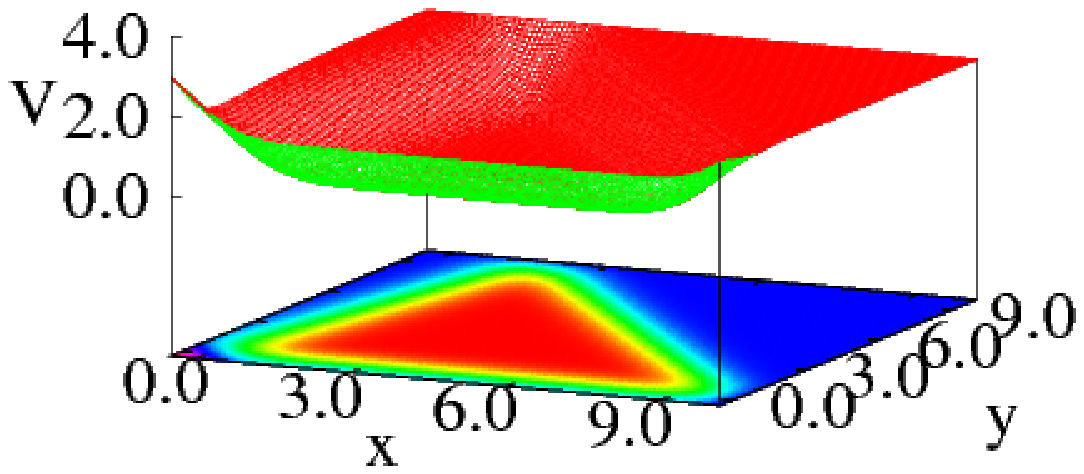}
 \includegraphics*[width=3.8cm,angle=0]{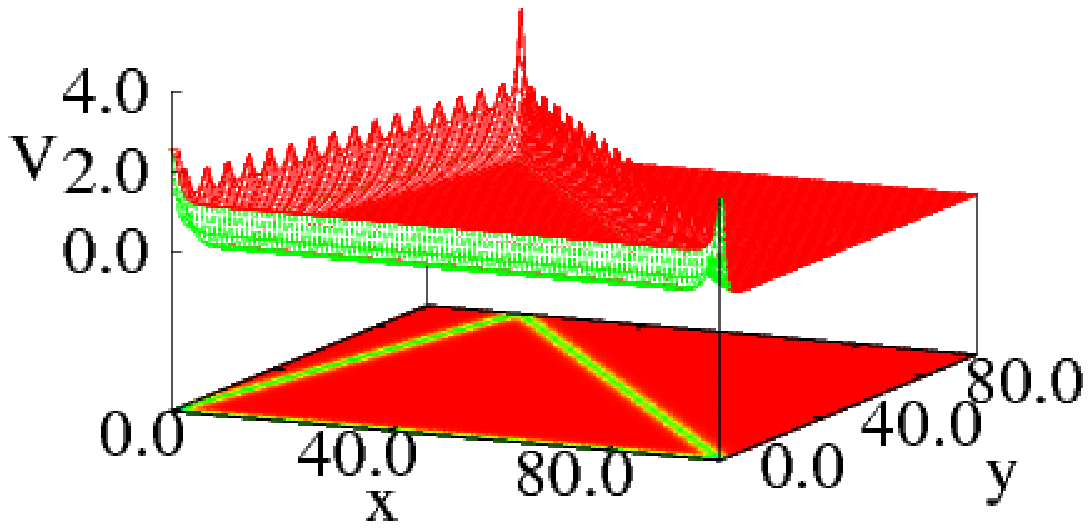}
 \includegraphics*[width=3.8cm,angle=0]{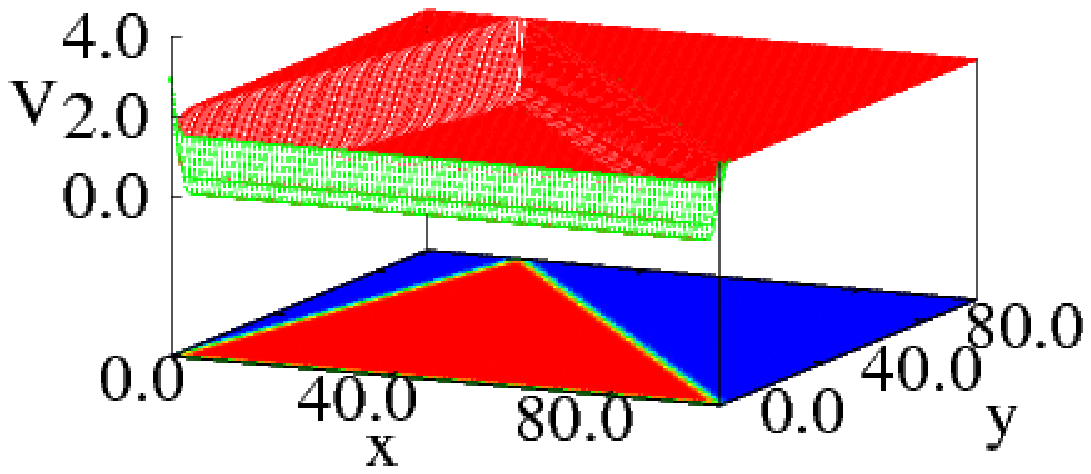}
 \end{center}
 \caption{(Color online) Potential energy from $  H_B^{exp}$ (left)
 and $H_B^{erf}$ (right) for the case of equal softness ratio and 
 $\gamma_{21}=5.0, \gamma_{31}=1.0$ and $\gamma_{32}=1.0/5.0$.
 Top row for the soft case $L/\sigma_{31}=10$ and right bottom 
row for  $L/\sigma_{31}=100$ approaching the hard-wall case.}
  \label{potentials5}
  \end{figure}

For unequal masses $\gamma_{21}=5.0, \gamma_{31}=1.0$ and 
$\gamma_{32}=1.0/5.0$
the internal angles of the soft billiard change, as can be seen in 
Fig.~\ref{potentials5}, where the softness ratio are constant, and 
top row shows results for $L/\sigma_{31}=10$ while bottom row
for  $L/\sigma_{31}=100$.  Still we observe that  $L/\sigma_{31}=100$
nicely approaches the hard-wall limit. 

To make a distinction between the hard-wall case from the billiard
system (\ref{Hdelta}), here we say that when  $L/\sigma_{31}$ 
increases, the quasi-hard wall limit is reached. In addition, even 
though we are using appropriate potentials, when 
$L/\sigma_{31}\gtrsim 100$ numerical difficults may appear 
because the potentials  becomes too steep. 

\subsection{The dynamics}

To analyze the dynamics related to the Hamiltonians from 
(\ref{three}) we look at the PPS for the total energy $E_B=2.0$,  
which is the higher energy possible before the particle becomes 
unbounded.
 \begin{figure}[htb]
 \unitlength 1mm
 \begin{center}
 \includegraphics*[width=4.2cm,angle=0]{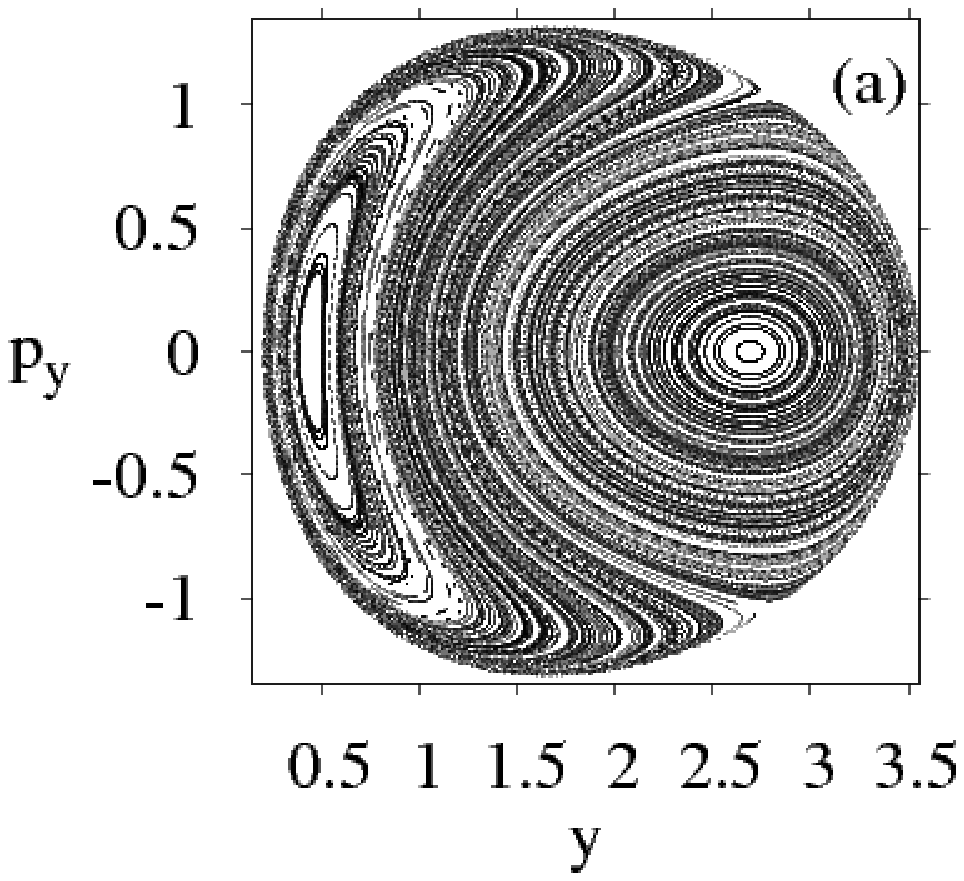}
 \includegraphics*[width=4.2cm,angle=0]{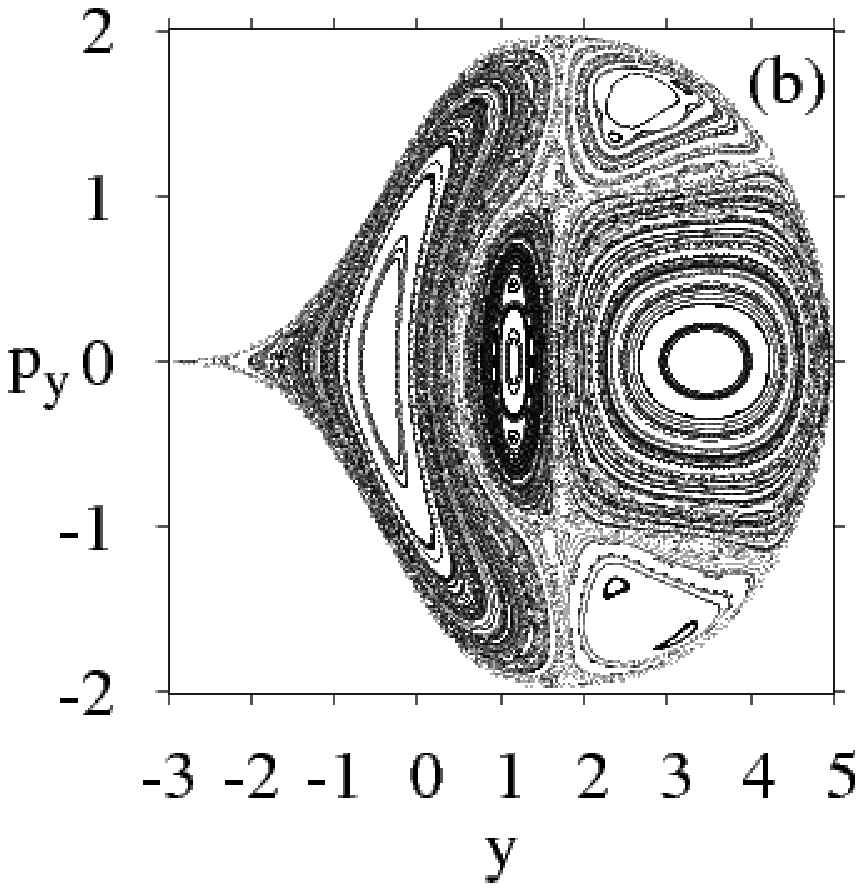}
 \includegraphics*[width=4.2cm,angle=0]{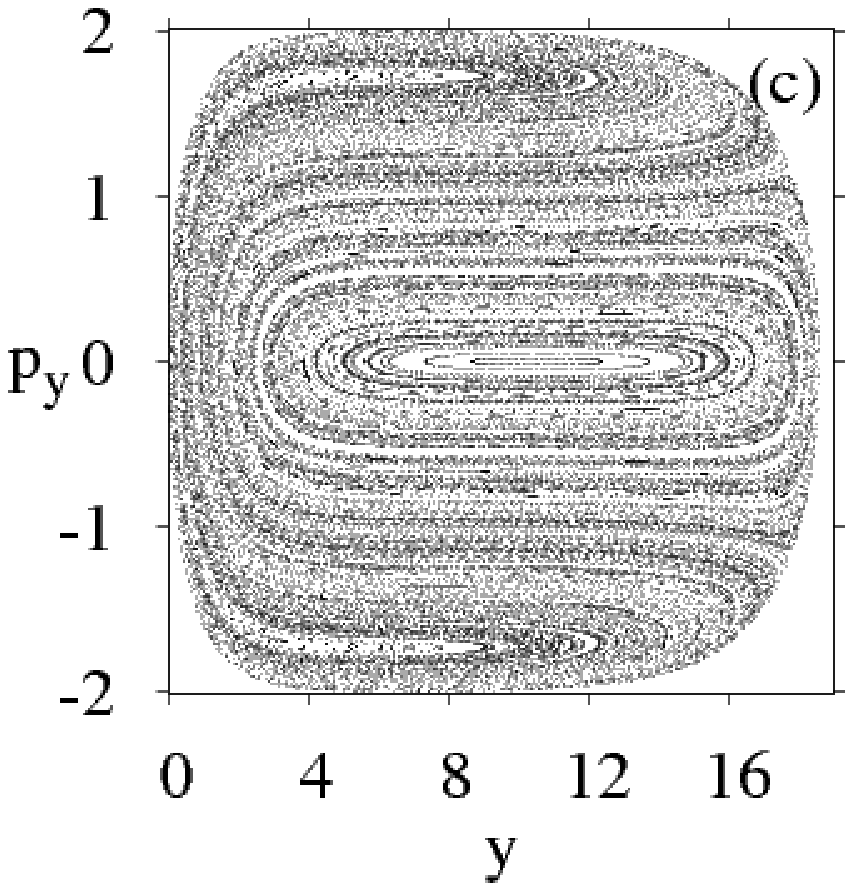}
 \includegraphics*[width=4.2cm,angle=0]{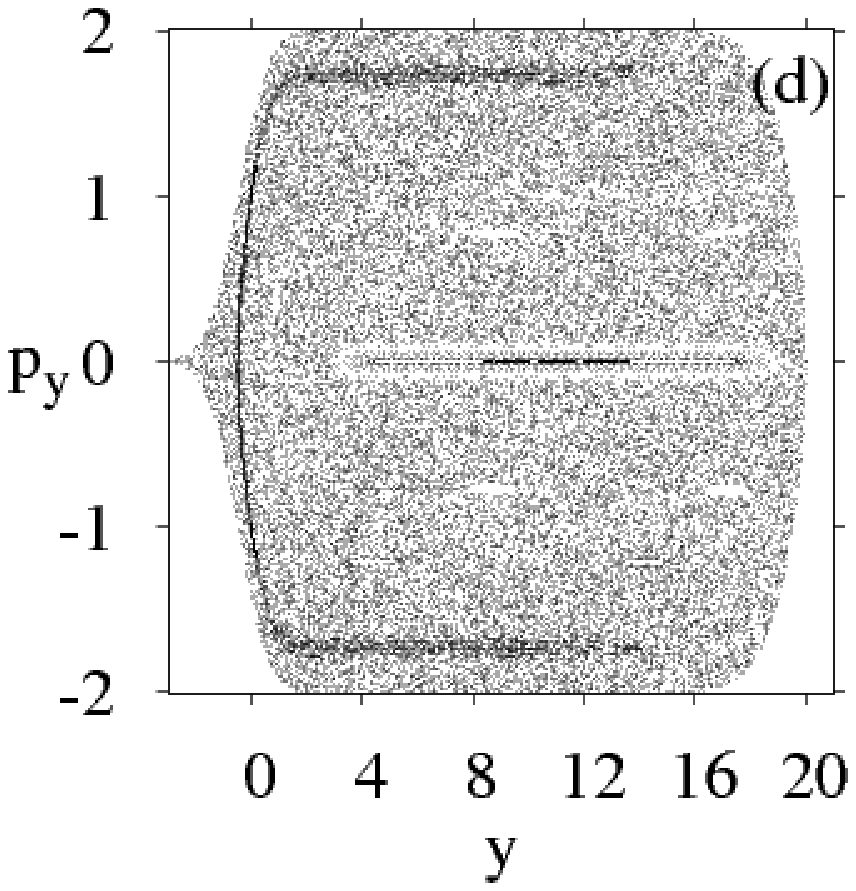}
 \end{center}
 \caption{PSSs for $H_B^{exp}$ (left) and $H_B^{erf}$ 
  (right) for the case of equal  masses and equal 
  softness ratio. Top PSSs are the soft case $L/\sigma_{31}=5.0$
  and bottom PSSs for the quasi-hard-wall case $L/\sigma_{31}=20$.}
  \label{PSS1}
  \end{figure}
We start showing the case of equal masses and equal softness 
ratios. Figure \ref{PSS1} compares the PSSs ($p_y,y,p_x>0$) for the 
system $ H_B^{exp}$ in Figs.~\ref{PSS1}(a),(c), $ H_B^{erf}$ in 
Figs.~\ref{PSS1}(b),(d). The PSSs from Figs.~\ref{PSS1}(a),(b) 
are for the soft case $L/\sigma_{31}=5.0$, while those from 
Figs.~\ref{PSS1}(c),(d) for the quasi-hard-wall $L/\sigma_{31}=20$. 
Figures \ref{PSS1}(a),(c) are related to the Toda potential 
discussed in section \ref{todae}, but now for $L/\sigma_{31}\ne 0$. 
The dynamics is regular, independent of softness parameter 
$L/\sigma_{31}$. In this quasi-hard integrable limit trajectories 
travel along one invariant irrational torus (straight horizontal
line) with $p_y>0$ ($p_y<0$), but when they reach the boundary 
with a very small softness, they invert the momentum and return 
along another invariant torus. The softness of the walls decide 
how fast this momentum inversion occurs, i.e.~for quasi-hard walls 
the inversion occurs relatively fast [Fig.~\ref{PSS1}(c)]. For the 
hard wall limit from (\ref{Hdelta}) it occurs instantaneously but
it does not change the torus. This is valid for these integrable 
systems. When the dynamics from 
$ H_B^{erf}$ is analyzed we see an almost regular motion 
in the soft walls case from Fig.~\ref{PSS1}(b). However, some 
hyperbolic points are present which generate a chaotic motion when
$L/\sigma_{31}$ increases. This is shown in Fig.~\ref{PSS1}(d) 
for the quasi-hard case $L/\sigma_{31}=20$, where the motion is 
chaotic with two sticky motions close to $p_y\sim 0.0,\pm 1.7$. 
The sticky motion is the consequences of the periodic motion which
occurs and can be clearly seen in the exponential quasi-hard limit 
for these points [Fig.~\ref{PSS1}(c)]. Besides these points the motion 
is totally chaotic. This nicely shows the numerical evidence that
the integrability condition found by the original paper from
\cite{toda70} is only valid when an {\it exponential} function is
used.  Therefore, even in the quasi-hard case 
$L/\sigma_{31}\to\infty$ small changes in the potential interaction 
function may drastically change the nature of the dynamics.
A chaotic motion was also observed (not shown) when the $V_{arc}(x)$ 
was used. It would be interesting if such integrability conditions 
using the exponential potentials could be applied to find another 
exact solutions in quantum scattering problems \cite{luz01,luz03}.

 \begin{figure}[htb]
 \unitlength 1mm
 \begin{center}
 \includegraphics*[width=4.2cm,angle=0]{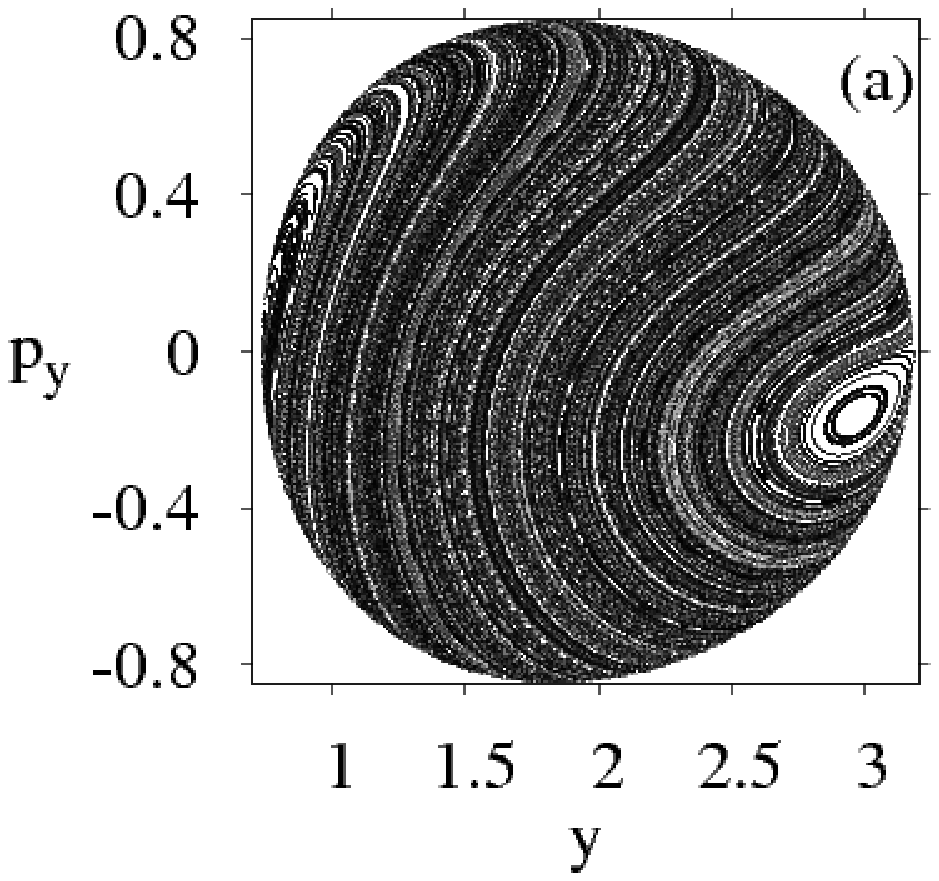}
 \includegraphics*[width=4.2cm,angle=0]{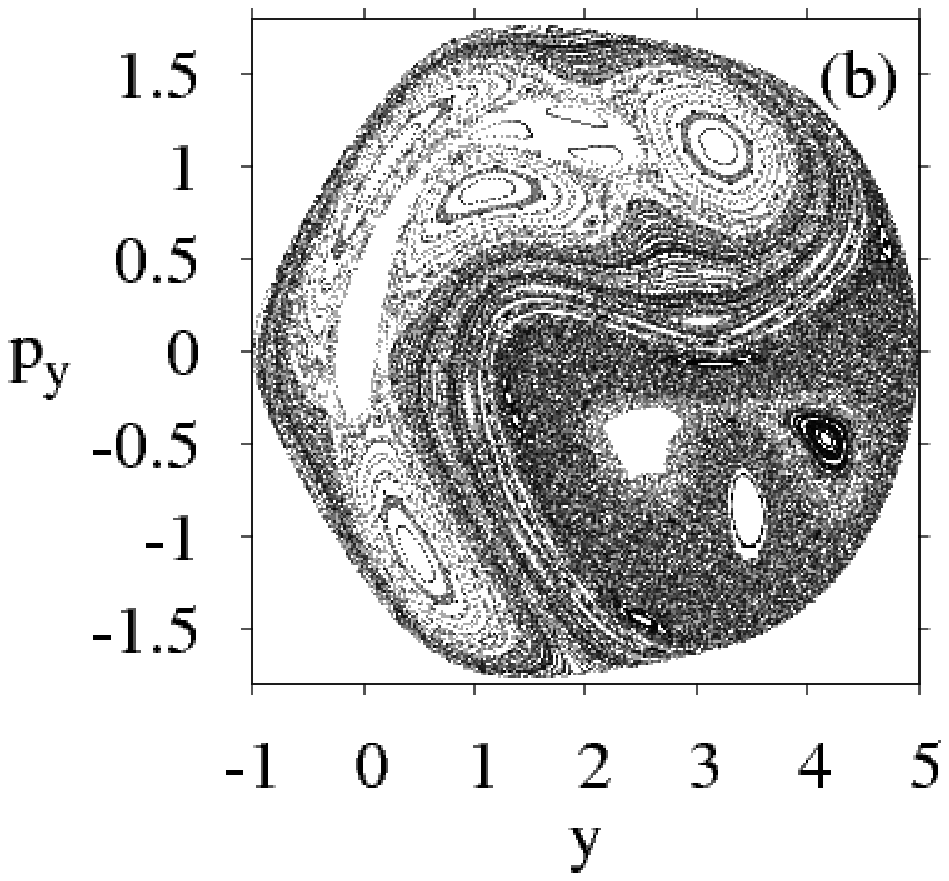}
 \includegraphics*[width=4.2cm,angle=0]{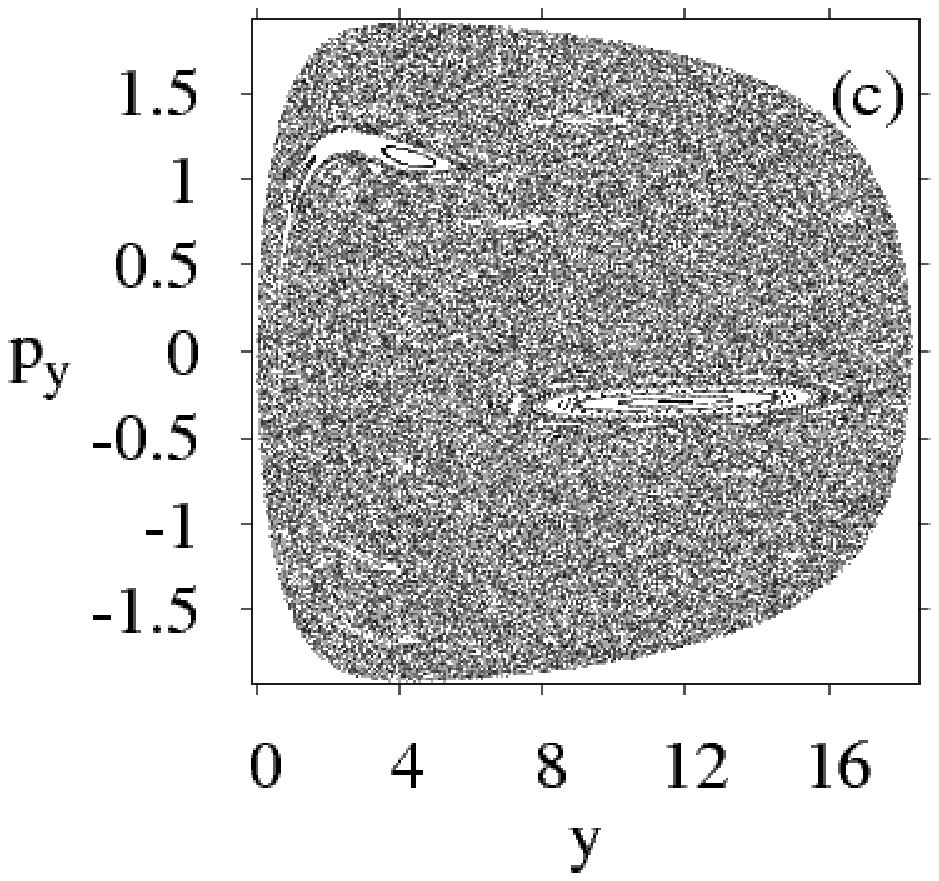}
 \includegraphics*[width=4.2cm,angle=0]{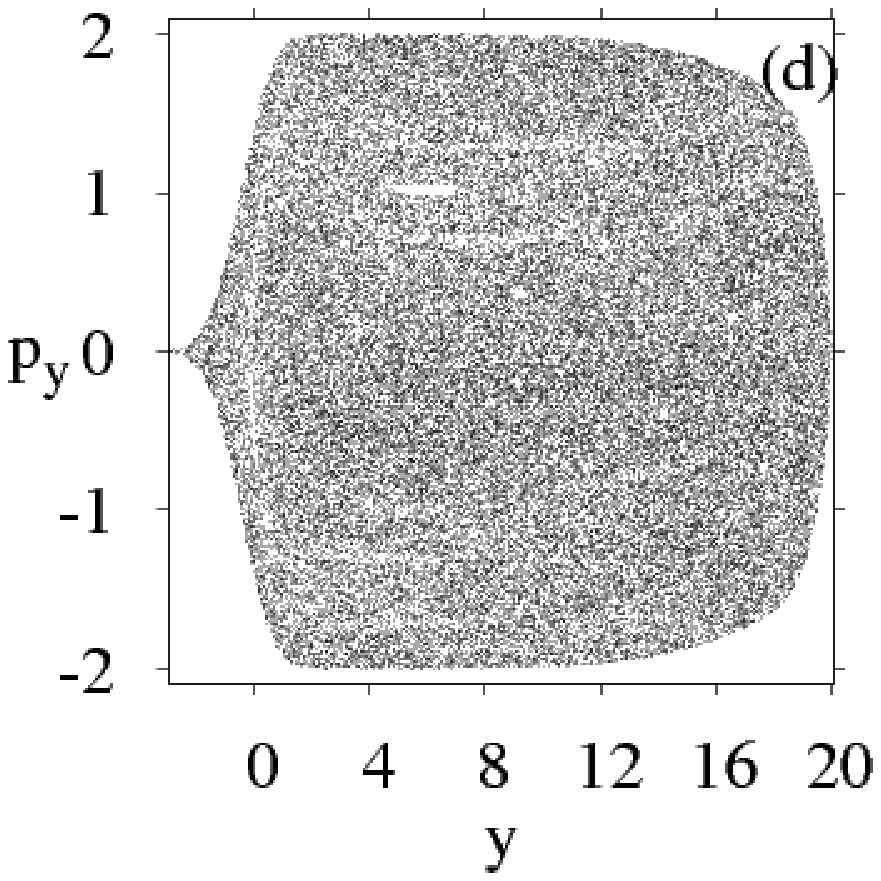}
 \end{center}
 \caption{PSS for the case of masses ratio  
 $\gamma_{21}=5.0, \gamma_{31}=1.0$ and $\gamma_{32}=1.0/5.0$
 and equal softness ratio  for $H_B^{exp}$ (left), $H_B^{erf}$ 
(right). Above is the soft case $L/\sigma_{31}=5$
  and below the hard-wall case $L/\sigma_{31}=20$.}
  \label{PSS2}
  \end{figure}
Figure \ref{PSS2} shows the dynamics when the masses ratio is
changed but for equal softness ratio.  The masses ratio are 
given by $\gamma_{21}=5.0, \gamma_{31}=1.0$ and 
$\gamma_{32}=1.0/5.0$. The left plots are for $H_B^{exp}$ and
the right plots for $H_B^{erf}$. Figs.~\ref{PSS2}(a) and (b) are
for the soft case  $L/\sigma_{31}=5.0$, showing that the motion
is almost regular for the exponential potential but mainly 
chaotic for the error potential.  Figs.~\ref{PSS2}(c) and (d) 
are for the quasi-hard wall limit $L/\sigma_{31}=20$. In both 
cases the dynamics is almost chaotic.

\section{Conclusions}
\label{conclusions}

The transition from soft to hard interaction between three particles 
on a frictionless ring is discussed.  It is a generalization 
of the two soft interacting  particles discussed in \cite{hercules08}.
Since usually hard interactions are modelled by $\delta$-potentials, 
for which the equations of motion are not well defined, we propose 
suitable potentials to study the soft to hard interaction transition, 
where the forces, not the potentials,
become $\delta$-functions in the limit of hard interactions. 
A scaled Hamiltonian is obtained which nicely shows this transition 
and gives general clues about the relevant properties of interacting 
particles: the dynamics depends only on the {\it masses ratio} 
between particles and {\it softness ratio} of the interaction, 
independent of the kind of interaction. Beside that any interaction 
potential with a parameter controlling the hard interacting limit can 
be used in this Hamiltonian which defines a {\it soft triangle} billiard 
inside which the whole dynamics occurs. The case of equal masses ratios 
is found to be integrable when a soft exponential interaction is 
assumed, like a Toda potential with a softness parameter. Equal masses
ratio with other interaction potentials, like the error function and 
the arctan function (not shown) generate a chaotic dynamics. We 
consider the dynamics of two suitable potentials, the exponential 
interaction with an additional softness parameter, a Toda like 
potential, and the error function. Results show that such potentials 
are appropriate and they do not present numerically divergencies when 
approaching the hard interaction limit.

\begin{acknowledgments}
The authors thank CNPq for financial support.
\end{acknowledgments}


\end{document}